\documentclass[12pt,twoside]{IEEEtran+}
\usepackage{amstex,amssymb}
\def\QED{\mbox{\rule[0pt]{1.5ex}{1.5ex}}}

\def\@begintheorem#1#2{\tmpitemindent\itemindent\topsep 0pt\rm\trivlist
    \item[\hskip \labelsep{\indent\it #1\ #2:}]\itemindent\tmpitemindent}
\def\@opargbegintheorem#1#2#3{\tmpitemindent\itemindent\topsep 0pt\rm \trivlist
    \item[\hskip\labelsep{\indent\it #1\ #2\ \rm(#3)}]\itemindent\tmpitemindent}
\def\@endtheorem{\endtrivlist\unskip}

\def\BibTeX{{\rm B\kern-.05em{\sc i\kern-.025em b}\kern-.08em
    T\kern-.1667em\lower.7ex\hbox{E}\kern-.125emX}}
\def\={\triangleq}                           %{\overset{\triangle}{=}}
\def\tr{\,{\rm tr}\,}
\newtheorem{thm}{Theorem}[section]
\newtheorem{defi}[thm]{Definition}
\newtheorem{rem}[thm]{Remark}
\newtheorem{cor}[thm]{Corollary}
\newtheorem{ex}[thm]{Example}
\newtheorem{lem}[thm]{Lemma}
\newtheorem{prop}[thm]{Proposition}
\makeindex

\begin{document}
%\addtocontents{toc}{\bigskip}

\title{\LARGE Quantum Channels and Simultaneous ID Coding}
\makeatletter 
\author{Peter L\"ober$^1$\footnote{$^1$Email: loeber@mathematik.uni-bielefeld.de}}
\maketitle
\makeatother 
\setcounter{footnote}{1}

{\footnotesize\centerline{(submitted to IEEE Trans.~Inform.~Theory)}}

\bigskip
\begin{abstract}
This paper is on identification of classical information by the use of 
quantum channels. We focus on simultaneous ID codes which use measurements
being useful to identify an arbitrary message. We give a direct and a
converse part of the appropriate coding theorem.
\end{abstract}

\section{Introduction}
%---------------------
\bigskip
Since 1948 when Shannon (\cite{sha48}) introduced information theory as a 
theory of communication there has been a quite big development in this field.
People realized that beyond Shannon's original models of communication there 
are further ones of interest for investigation.
One of those is the theory of identification (ID) via channels introduced in 
1989 by Ahlswede/Dueck (\cite{ahl89}). Here the
receiver is not interested in the exact message. He only wants to know if 
the sent message is a special one he is interested in.
The authors gave a nice proof that under these constraints there are codes 
with doubly exponential size in the block length of the codewords.
Even though this was a big surprise, it is (at least technically) much 
harder to give a satisfying converse to this coding theorem.
A strong converse was given by Han/Verd\'u (\cite{han92}) in 1992. (The 
discussion of other interesting code models may be found in \cite{ahl97}.)
\smallskip\\
The goal of this paper is to study the ID scheme for the case of the
information being transfered by a quantum channel. We give a 
coding theorem and a strong converse theorem for ID coding schemes that use
measurements which may be used to identify every message. This means that the
measurement the receiver has to perform to access the information
is not allowed to depend on the message he's actually interested in.\\
Investigation in quantum channels started in the 1960's (see \cite{hol98} for 
a list of references), leading to Holevo's famous upper bound (\cite{hol73})
which implies immediately a weak converse for the transmission problem of the
memoryless quantum channel.
Even though it was undoubted that the appropriate coding theorem holds it was 
not before 1996 when people were able to prove this direct part of the 
coding theorem (\cite{hau96}, \cite{hol96}, \cite{sch97}). Today it's also 
known that the strong converse holds (\cite{oga98}, \cite{win98}), and that
things work for non-stationary quantum channels, too (\cite{win99}).\\
The contents of this paper essentially coincides with the contents of my
preprint \cite{loe99}.\\

%-----------------------------------------------------------------------------

%\clearpage\pagestyle{empty}\cleardoublepage\pagestyle{plain}
\section{Basic Definitions and Main Results}
%-------------------------------------------
\bigskip
\begin{defi}\rm
Let $A=\{1,...,a\}$ be a finite set and let $\cal H$ be a finite dimensional 
(complex) Hilbert space with $S({\cal H})\subset{\cal L(H)}^*$ its
corresponding \it state space \index{state space $S({\cal H})$}\rm
\footnote{${\cal L(H)}$ denotes the space of linear operators on ${\cal H}$.}
(positive and unity preserving linear forms 
on $\cal L(H)$). A \it quantum channel 
\index{quantum channel}\rm ${\bf W}=(W^n)_{n\in\mathbb N}$ is 
a sequence of maps 
$$ W^n:A^n\to S({\cal H}^{\otimes n}) \qquad x^n\mapsto W^n_{x^n} \ .$$
We call it \it memoryless \index{memoryless quantum channel}\rm if
$W^n_{x^n} = W^1_{x_1}\otimes\ldots\otimes W^1_{x_n}$ for all
$x^n=(x_1,\ldots,x_n)\in A^n$.\\
\end{defi}

To access the (classical) information of a quantum state we have to perform
a measurement on the output space:\\

\begin{defi}\rm
Let $\cal H$ be a finite dimensional (complex) Hilbert space. A \it POM 
(positive operator measurement) on $\cal H$ 
\index{positive operator measurement (POM)}\rm is a tuple
$D=(D_i)_{i=1,\ldots,N}$ of non-negative operators $D_i$ on $\cal H$ such that
$\sum_{i=1}^N D_i = {\bf 1}_{\cal H}$. Here ${\bf 1}_{\cal H}$ denotes
the unity operator on $\cal H$.\\
\end{defi}

\begin{rem}\rm
A POM is a kind of resolution of unity. Its practical interpretation is the
following: Given a state $\sigma\in S({\cal H})$ the probability that the
result of measurement $D$ will be $i$ is $\sigma(D_i)$.\\
\end{rem}

\begin{ex}\rm
Let $\cal H$ be a finite dimensional (complex) Hilbert space with orthonormal 
basis $(\psi_i)_{i=1,\ldots,N}$. Let $D_i$ be the projector on $\psi_i$. Then
the tupel $D=(D_i)_{i=1,\ldots,N}$ is a POM, a so called \it von Neumann
measurement.\index{von Neumann measurement}\rm\\
It can be shown that any POM on $\cal H$ may be interpreted as a von Neumann 
measurement on an (occasionally) larger system. This is known as Naimark's 
theorem. (For a rigorous formulation and proof see p.~65 of \cite{hol82}.)
\end{ex}

We start with the definitions for the transmission problem because we will
present our results on ID capacities relatively to those for transmission:\\

\begin{defi}\rm
An $(n,M,\varepsilon)$ \it Q code 
\index{Q code}\rm is defined to be a set of pairs 
$\{(c_m,E_m) : m=1,\ldots,M\}$ with $c_i\in A^n$ and $E\=(E_m)_{m=1,\ldots,M}$ 
a POM on ${\cal H}^{\otimes n}$ such that ($\forall\,m=1,\ldots,M$):
$$ W^n_{c_m}(E_m) \ge 1 - \varepsilon\ .$$\smallskip
\end{defi}

\begin{defi}\rm\label{TrmCap}
Given a quantum channel ${\bf W}$ denote the maximum $M$ such that there is an 
$(n,M,\varepsilon)$ Q code by $M(n,\varepsilon)$. For $0<\varepsilon<1$ we 
introduce a \it pessimistic \rm and an \it optimistic $\varepsilon$-error 
capacity 
\index{capacity}\index{pessimistic capacity}\index{optimistic capacity}\rm by:
$$  C(\varepsilon) \= 
   \liminf_{n\to\infty}\frac{\log M(n,\varepsilon)}{n} \qquad {\rm and} \qquad
   \bar{C}(\varepsilon) \= 
   \limsup_{n\to\infty}\frac{\log M(n,\varepsilon)}{n}\ .$$
We define the following four \it capacities:\rm
$$ C_0 \= \inf_{\varepsilon>0} C(\varepsilon)\ , \quad
   C_1 \= \sup_{\varepsilon<1} C(\varepsilon)\ , \quad
   \bar{C}_0 \= \inf_{\varepsilon>0} \bar{C}(\varepsilon)\ , \quad
   \bar{C}_1 \= \sup_{\varepsilon<1} \bar{C}(\varepsilon)\ .$$\smallskip
\end{defi}

We mention that (obviously) $C_0 \le C_1, \bar{C}_0 \le \bar{C}_1$, 
and turn directly to the ID code model:\\
Whereas for Shannon's transmission problem the receiver wants to know exactly
which message was sent, in the ID model the receiver only wants to check
if it is some (fixed) message $i$. The sender (of course) does not know
which message the receiver is interested in. The canonical model for 
a quantum version of the ID code model is  the following (cf. \cite{ahl89} 
for further motivation and examples):\\

\begin{defi}\rm
A \it (randomized) $(n,N,\lambda_1,\lambda_2)$ Q-ID code 
\index{Q-ID code}\index{randomized Q-ID code}\rm is a set of pairs 
$\{(P_i,D_i):i=1,\ldots,N\}$ where the $P_i$s are probability distributions on
$A^n$ and the $D_i$s, ${\bf 0} \le D_i\le{\bf 1}$, denote operators on 
${\cal H}^{\otimes n}$ such that for all $i,j= 1,\ldots,N$ with $i\not= j$:
$$ P_iW^n(D_i) \ge 1-\lambda_1 \quad{\rm and}\quad
   P_iW^n(D_j) \le \lambda_2 \ .$$
Here and in the following we use $\ P_iW^n 
\= \sum_{x^n\in A^n} P_i(x^n)W^n_{x^n}\,\in S({\cal H}^{\otimes n})$
as an abbreviation.\\
\end{defi}

We draw attention to the fact that we use random encoding, which means that
a message is represented by a probability distribution on the possible
codewords and (in general) not by a single codeword (cf.\cite{ahl89}).\par
It is important to realize that for the Q-ID code model (above) the $D_i$s are
not 
supposed to form a POM. Each $D_i$ for itself (together with ${\bf 1}\!-\!D_i$)
could be thought of as a POM, namely as the POM the receiver performs asking 
for message $i$. Therefore Q-ID codes have a remarkable property that is
different from the classical (ID) case: The receiver can't in general use the 
same received state to ask for two different messages $i$ and $j$ because 
asking for message $i$ includes a measurement on this state. To overcome this 
problem we formulate a second code model for which there has to be one single 
(simultaneous) measurement which allows to identify every message at the same 
time. This model is also valid if the one who performs the measurement is not 
the (final) receiver himself and doesn't also know in which
message this receiver is interested in.\\

\begin{defi}\rm\label{SimDef}
A Q-ID code $\{(P_i,D_i):1\le i\le N\}$ is called \it simultaneous 
\index{simultaneous Q-ID code}\rm if there is a POM 
$(E_m)_{m\in{\cal M}}$ and there are subsets 
${\cal A}_i\subseteq{\cal M}\ (1\le i\le N)$
such that $D_i = \sum_{m\in{\cal A}_i} E_m$.\\
\end{defi}

$(E_m)_m$ should be viewed as a common refinement of the resolutions of unity
$(D_i,{\bf 1}\!-\!D_i)_{i=1,\ldots,N}$.\medskip\\
We remark at this point that most examples (!) of ID coding schemes require 
simultaneous ID codes because their real implementation would consist of many 
receivers (at one time). This holds for the examples given in 
\cite{ahl89}. On the other hand this is not always the case, e.g.~if both, 
sender and receiver, have a (possibly different) text and they want to check if
it is the same one, using an ID code. Here really is only one receiver asking
only one question.\footnote{This example is taken from \cite{klei99}.}\\

\begin{defi}\rm
Denote the maximum $N$ such that there is an 
$(n,N,\lambda_1,\lambda_2)$ Q-ID code by $N(n,\lambda_1,\lambda_2)$. The
\it ID capacities 
\index{ID capacity}\rm are defined as follows (cf.~\cite{ahl89}):
\begin{eqnarray}
C(\lambda_1,\lambda_2) &\=& \liminf_{n\to\infty} 
      {\log\log N(n,\lambda_1,\lambda_2) \over n}, \quad {\rm and}\nonumber\\
   \bar{C}(\lambda_1,\lambda_2) &\=& \limsup_{n\to\infty} 
      {\log\log N(n,\lambda_1,\lambda_2) \over n}\ .\nonumber
\end{eqnarray}
\end{defi}

Accordingly, we define $N^{sim}(n,\lambda_1,\lambda_2)$,
$C^{sim}(\lambda_1,\lambda_2)$ and $\bar{C}^{sim}(\lambda_1,\lambda_2)$
for the smaller class of simultaneous Q-ID codes,
following the same lines.\\

\begin{rem}\rm
$N(n,\lambda_1,\lambda_2) \ge N^{sim}(n,\lambda_1,\lambda_2)$,
$C(\lambda_1,\lambda_2) \ge C^{sim}(\lambda_1,\lambda_2)$, and
$\bar C(\lambda_1,\lambda_2) \ge \bar C^{sim}(\lambda_1,\lambda_2)$.\\
\end{rem}

We are now able to state the main results of this paper:\\

\begin{thm}\rm\label{DirSim}\index{Q-ID coding theorem}
Let $\lambda_1,\lambda_2>0$. Then
$$ C^{sim}(\lambda_1,\lambda_2) \ge C_0 \qquad{\rm and} \qquad
   \bar C^{sim}(\lambda_1,\lambda_2) \ge \bar C~\!\!_0 \ . $$\smallskip
\end{thm}

\begin{cor}\rm\label{DirQID}
Let $\lambda_1,\lambda_2>0$. Then
$$ C(\lambda_1,\lambda_2) \ge C_0 \qquad {\rm and} \qquad
   \bar C(\lambda_1,\lambda_2) \ge \bar C~\!\!_0 \ . $$\smallskip
\end{cor}

\begin{thm}\rm\label{ConSim}\index{converse Q-ID coding theorem}
Let $\lambda_1+\lambda_2 < 1$. Then
$$ C^{sim}(\lambda_1,\lambda_2) \le C_1 \qquad{\rm and} \qquad
   \bar C^{sim}(\lambda_1,\lambda_2) \le \bar C~\!\!_1 \ . $$\smallskip
\end{thm}

\begin{cor}\rm\label{MemSim}
Since it is known for memoryless (!) quantum channels that all the four 
transmission capacities of Definition \ref{TrmCap} are equal to one constant
$C$, we have for all $\lambda_1,\lambda_2>0$ with $\lambda_1+\lambda_2 < 1$
$$ C^{sim}(\lambda_1,\lambda_2) = \bar C^{sim}(\lambda_1,\lambda_2) = C\ .$$
Here as usual, $C$ fulfills the formula
$$ C = \max_{P\text{ \rm PD on }A} 
       \left(H(\hat{PW})-\sum_{x\in A}P(x)H(\hat W_x)\right)\ ,$$
with $H(\hat W_x)=-\tr(\hat W_x\cdot\log \hat W_x)$, where for a state 
$\sigma\in S({\cal H})$ we wrote $\hat\sigma\in{\cal L(H)}$ for
the uniquely defined operator with $\sigma = \tr(\hat\sigma\,\cdot\,)$.
(See \cite{hau96}, or for general input states \cite{hol96} or \cite{sch97},
for a proof of $C_0 \ge C$, and \cite{oga98} or \cite{win98} for
$\bar C_1 \le C$.) Of course, our theorems apply to other quantum channels, 
too, e.g. to the non-stationary quantum channels (cf. \cite{win99}).\\
\end{cor}

We shall prove Coding Theorem \ref{DirSim} in the next section.
At the end of Section 4 there is a proof of the Converse Theorem \ref{ConSim}.
(This proof will be completed by a theorem we prove in Section 5.)\\

\begin{rem}\rm
It is an open question whether (the analogue of) Converse Theorem \ref{ConSim}
holds
in the general (non-simultaneous) case, too (cf.~also Remark \ref{TrivCon}).\\
\end{rem}

%\begin{thm}\rm\label{SimQID}
%For all $0 < \lambda_1+\lambda_2 < 1$ it holds 
%$$C^{sim}(\lambda_1,\lambda_2)=C^{Sh}\ .$$ 
%Here $C^{Sh}$ denotes the Shannon$\,$-$\,$von$\,$Neumann capacity 
%$$ C^{Sh}=
%   \max_{P\text{ \rm PD on }A} \left(H(PW)-\sum_{x\in A}P(x)H(W_x)\right) $$
%of the channel $W$ with $H(W_x)=-\tr(W_x\cdot\log W_x)$.
%\end{thm}
%
%\begin{cor}\rm\label{QIDcoding}
%For all $0 < \lambda_1+\lambda_2 < 1$ it holds 
%$$C(\lambda_1,\lambda_2) \ge C^{Sh}\ .$$ 
%\end{cor}
%
%-----------------------------------------------------------------------------
\bigskip
%\clearpage\pagestyle{empty}\cleardoublepage\pagestyle{plain}
\section{Direct Part of Simultaneous Q-ID Coding}
%-----------------------------------------------
\bigskip
For this section we were fortunately able to follow \cite{ahl89} directly.
We formulate a lemma that is up to slight modifications nothing else but the 
main proposition used in that paper:\\

\begin{lem}\rm\label{GilbLem}
Let ${\cal M}$ be a finite set of cardinality $M$ and let $\lambda\in(0,1)$.
Let $\varepsilon>0$ be so small such that 
$\lambda\log_2(\frac{1}{\varepsilon}-1) > 2$. Then there are at least 
$N \ge \frac{1}{M}2^{\lfloor\varepsilon M\rfloor}$ subsets 
${\cal A}_1,\ldots,{\cal A}_N\subset {\cal M}$, each of cardinality
$\lfloor\varepsilon M\rfloor$, such that the cardinalities of the
pairwise intersections fulfill
$$ |{\cal A}_i \cap {\cal A}_j| < \lambda\lfloor\varepsilon M\rfloor
        \qquad\qquad \forall\ i,j=1,\ldots,N \ (i\not= j)\ .$$\smallskip
\end{lem}

\begin{proof}
Let $N$ be the maximum number such that there is a family of (distinct) subsets
${\cal A}_1,\ldots,{\cal A}_N\subset {\cal M}$ with the desired
properties. Let $a\=\lfloor\varepsilon M\rfloor$. For each $i=1,\ldots,N$ we 
count the number of subsets ${\cal A}\subset {\cal M}$ with cardinality $a$ but
$|{\cal A}_i \cap {\cal A}| \ge \lambda a$. This number is
$$ \sum_{k=\lceil\lambda a\rceil}^a {M-a\choose a-k}{a\choose k}
   \ \le\ \sum_{k=\lceil\lambda a\rceil}^a {M\choose a-k}\, 2^a
   \ \le\ a{M\choose a-\lceil\lambda a\rceil}\, 2^a\ .$$
Defining $S\=a{M\choose a-\lceil\lambda a\rceil}\, 2^a$ we could add another
set to our family of subsets if ${M\choose a} > N\!\cdot\! S$. Therefore:
$$ N \ \ge\ \frac{1}{S} {M\choose a} 
     \ \ge\ \frac{1}{a}\, 2^{-a}\, (\underbrace{
   \frac{M-a}{a}}_{\ge\frac{1}{\varepsilon}-1})^{\lceil\lambda a\rceil}
     \ \ge\ 
\frac{1}{M}\, 2^{-a}\, 2^{\lceil\lambda a\rceil\log_2(\frac{1}{\varepsilon}-1)}
     \ \ge\ \frac{1}{M}\, 2^a \ .$$\end{proof}\bigskip

%We will use (cf.\cite{ahl89}) the coding theorem for transmission as it
%is proved in\cite{hau96} or for general input states in\cite{hol96} or
%\cite{sch97}:
%
%\begin{thm}\rm\label{TRMcode}
%Let $\delta,\lambda > 0$. There is a positive number $n_0$ such that
%for all $n\ge n_0$ there exists some transmission code
%$\{(c_m,E_m) : m=1,\ldots,M \}$ with $c_m\in A^n$ and the $E_m$ form a
%POM on ${\cal H}^{\otimes n}$ such that $M\ge 2^{(C^{Sh}-\delta)n}$ and
%$$ \tr(E_m\cdot W^n_{c_m})\ \ge\ 1-\lambda \qquad \forall\ m=1,\ldots,M\ .$$
%\end{thm}

Coding Theorem \ref{DirSim} is an immediate consequence of the following 
proposition (see also Remark \ref{DirSup}).\\
%$(\,C(\lambda_1,\lambda_2)\ge C^{sim}(\lambda_1,\lambda_2)\ge C^{Sh}\,)$.

\begin{prop}\rm
Let $\lambda_1,\lambda_2,\delta>0$, let
$\lambda\=\min(\lambda_1, ^{\lambda_2}\!\!/_2)$, and let $\varepsilon>0$ be so
small that $\lambda\log(\frac{1}{\varepsilon}-1) > 2$. Then there is a positive
number $n_0$ such that for all $n\ge n_0$ there exists some simultaneous 
$(n,N,\lambda_1,\lambda_2)$ Q-ID code $\{(P_i,D_i) : i=1,\ldots,N\}$ with 
$N\ge 2^{\,\lfloor\varepsilon 2^{(C_0-\delta)n}\rfloor - n}$.
\end{prop}

\begin{proof}
By definition of $C_0$ there is an $(n,M,\lambda)$ Q code
${\cal C}=\{(c_m,E_m) : m=1,\ldots,M \}$ with $M\ge 2^{(C_0-\delta)n}$ if only
$n$ is large enough. Using ${\cal M}=\{1,\ldots,M\}$ as ground set, 
Lemma \ref{GilbLem} provides us with subsets 
${\cal A}_1,\ldots,{\cal A}_N\subset{\cal M}$ of cardinality
$\lfloor\varepsilon M\rfloor$ with pairwise intersections smaller than
$\lambda\lfloor\varepsilon M\rfloor$. Here we have for the number $N$ of 
those sets:
$$ N \ge \frac{1}{M}2^{\lfloor\varepsilon M\rfloor}
\underset{n\gg 1}\ge 2^{\,\lfloor\varepsilon 2^{(C_0-\delta)n}\rfloor - n}\ .$$
We construct a simultaneous Q-ID code $\{(P_i,D_i):i=1,\ldots,N\}$ by taking 
as $P_i$ the uniform distribution on ${\cal C}_i \= \{c_m : m\in{\cal A}_i\}$
and as $D_i$ the sum of the corresponding $E_m$s:
$$ P_i(x^n) \= 
   \begin{cases}
   \frac{1}{\lfloor\varepsilon M\rfloor} & \text{\rm if $x^n\in{\cal C}_i$,}\cr
   \hspace{9pt} 0 & else,
    \end{cases}    \qquad {\rm and} \qquad
   D_i \= \sum_{m\in{\cal A}_i} E_m \qquad
   (i=1,\ldots,N) \ .$$
It's now straight forwards to calculate that the errors are small:
\begin{eqnarray}
 P_iW^n(D_i)
%   =  \tr(\sum_{m\in{\cal A}_i}E_m\cdot\sum_{x^n\in A^n}P_i(x^n)W^n_{x^n})
  &=& \frac{1}{\lfloor\varepsilon M\rfloor}       \nonumber
       \sum_{m\in{\cal A}_i}\sum_{m'\in{\cal A}_i} W^n_{c_{m'}}(E_m)\\
 &\ge& \frac{1}{\lfloor\varepsilon M\rfloor}
      \sum_{m\in{\cal A}_i} W^n_{c_m}(E_m)
 \ \ge\ 1-\lambda\ \ge\ 1-\lambda_1\ , \nonumber
\end{eqnarray}
and for $i\not= j$:
\begin{eqnarray}
&\ &\hspace{-1.2cm} P_iW^n(D_j)
  \ =\ \frac{1}{\lfloor\varepsilon M\rfloor}
      \sum_{m\in{\cal A}_j}\sum_{m'\in{\cal A}_i} W^n_{c_{m'}}(E_m)
   \nonumber\\
   & = & \frac{1}{\lfloor\varepsilon M\rfloor} \sum_{m\in{\cal A}_j} 
  \left(\sum_{m'\in{\cal A}_i\cap{\cal A}_j} W^n_{c_{m'}}(E_m)\
 +\sum_{m'\in{\cal A}_i\backslash{\cal A}_j} W^n_{c_{m'}}(E_m)\right)
   \nonumber\\
   & = & \frac{1}{\lfloor\varepsilon M\rfloor} 
         \sum_{m'\in{\cal A}_i\cap{\cal A}_j}
         \underbrace{W^n_{c_{m'}}(\sum_{m\in{\cal A}_j}E_m)}_{\le\ 1}
         \nonumber\\    
   & &\hspace{1cm}  +\quad\frac{1}{\lfloor\varepsilon M\rfloor} 
         \sum_{m'\in{\cal A}_i\backslash{\cal A}_j}
    \underbrace{W^n_{c_{m'}}(\sum_{m\in{\cal A}_j}E_m)}_{\le\ \lambda}
   \nonumber\\
  & \le & \frac{1}{\lfloor\varepsilon M\rfloor} 
          \cdot\lambda\lfloor\varepsilon M\rfloor 
  \ +\ \frac{1}{\lfloor\varepsilon M\rfloor}
       \cdot\lfloor\varepsilon M\rfloor\cdot\lambda
  \quad=\quad 2\,\lambda \quad\le\quad \lambda_2\ .\nonumber
\end{eqnarray}\end{proof}\bigskip

\begin{rem}\rm\label{DirSup}
It is obvious that the previous proposition still holds for (at least) an
infinite number of $n\in{\mathbb N}$ if we replace $C_0$
by (the possibly larger) $\bar C_0$. This shows that 
$\bar C^{sim}(\lambda_1,\lambda_2) \ge \bar C_0$.         %\\
Moreover it's clear that the slightly stronger statements
$C^{sim}(\lambda_1,\lambda_2) \ge C(\lambda)$ and
$\bar C^{sim}(\lambda_1,\lambda_2) \ge \bar C(\lambda)$ hold for 
$\lambda_1,\lambda_2>0$ and $\lambda\=\min(\lambda_1, ^{\lambda_2}\!\!/_2)$.\\
\end{rem}

%----------------------------------------------------------------------------

%\clearpage\pagestyle{empty}\cleardoublepage\pagestyle{plain}
\section[A Resolvability Theory for Quantum Channels]{A Resolvability Theory for Quantum\\ Channels}
%---------------------------------------------------
\bigskip
In this section we develop a resolvability theory for quantum channels. This 
theory arises quite naturally from those for classical channels (cf.
\cite{han93}). Speaking very loosely it concerns the following question: 
Say that two input distributions are similar if the variational distance
$d_1(P,Q)\=\sum_{x\in A} |P(x)-Q(x)|$ of the corresponding output (!)
distributions is small. How small may be a set of input distributions
under the constraint that it represents all input distributions up to
similarity?\par
As we focus on simultaneous ID coding we will have to make our definitions
dependent on a fixed measurement. Recall that there are certainly very 
useless measurements (e.g.~trivial ones), which means that applications
of the results should only be of interest if one uses special (non-trivial)
measurements (e.g.~the underlying measurement of a ``good'' simultaneous Q-ID
code). Like above we will prefer the notion of 
probability distributions instead of random variables.\\

%Of course there is a difference whether we have at the output a fixed 
%measurement or we allow an arbitrary measurement, e.g. one that maximizes
%$d_1$, reflecting simultaneous Q-ID coding and general Q-ID coding,
%respectively. In the case of the fixed measurement we have to be aware that
%there are certainly very useless measurements (e.g. trivial ones). That's the 
%reason why at the end there will be some maximization (to get a useful result).

\begin{defi}\rm
Let $A$ be a finite set and let ${\cal H}$ be a finite dimensional Hilbert 
space. A \it process \index{process}\rm 
${\bf P}$ on $A$ is a sequence ${\bf P}=(P^n)_{n\in\mathbb N}$ 
with $P^n$ a probability distribution on $A^n$, a 
\it measurement process \index{measurement process}\rm ${\bf E}$ on 
${\cal H}$ is a sequence ${\bf E}=(E^n)_{n\in\mathbb N}$ with $E^n$ a POM on 
${\cal H}^{\otimes n}$. We call $({\bf P},{\bf E})$ a 
\it pair of processes.\index{pair of processes}\rm\\
\end{defi}

%\begin{defi}\rm
%Let ${\cal H}$ a finite dimensional Hilbert space. A measurement process
%${\bf E}$ on ${\cal H}$ is a sequence ${\bf E}=(E^n)_{n\in\mathbb N}$ with $E^n$ 
%a POM on ${\cal H}^{\otimes n}$.
%\end{defi}

\begin{defi}\rm
Let $P$ be a probability distribution on $A$. $P$ is \it $M$-type 
\index{type}\rm ($M\in\mathbb N$) if $P(x) \in
\{0,\frac{1}{M},\frac{2}{M},\ldots,\frac{M-1}{M},1\}$ for all $x\in A$.\\
\end{defi}

\begin{rem}\rm
Obviously, the number of different $M$-type distributions is upper bounded by 
$|A|^M$.\\
\end{rem}

\begin{defi}\rm
Let $P$ be a probability distribution on $A$. We call
$$ R(P) \= \min\{ M\in{\mathbb N} : \text{$P$ is $M$-type} \} $$
the \it resolution 
\index{resolution}\rm of $P$. (Let $R(P)\=\infty$ if $P$ isn't $M$-type 
for all $M\in\mathbb N$.)\\
\end{defi}

\begin{defi}\rm\label{distance}
Let $\sigma\in S({\cal H})$ be a state and let $E=(E_m)_{m=1,\ldots,M}$
be a POM on a finite dimensional Hilbert space ${\cal H}$. 
This induces a probability distribution 
$\sigma(E)$ on $\{1,\ldots,M\}$ with $\sigma(E)(m)\=\sigma(E_m)$. Given a
second state $\rho\in S({\cal H})$ let
$$ d_E(\rho,\sigma) \= d_1(\rho(E),\sigma(E)) \ .$$\\
\end{defi}

%\begin{rem}\rm
%It holds 
%$$ d(\rho,\sigma) = \max_{E\ {\rm POM\ on}\ {\cal H}} d_E(\rho,\sigma)\ .$$
%\end{rem}
%
%\begin{proof}
%Let $\{v_j\}$ an ONB of eigenvectors of $A:=\rho-\sigma$.
%\begin{eqnarray}
%d_E(\rho,\sigma) &=& \sum_i |\tr(E_iA)|
%    = \sum_i |\sum_{j,k} <v_j|E_i|v_k><v_k|A|v_j>|
%    = \sum_i |\sum_j <v_j|E_i|v_j><v_j|A|v_j>| \nonumber\cr
%&\le& \sum_{i,j} <v_j|E_i|v_j>|<v_j|A|v_j>|
%    = \sum_j |<v_j|A|v_j>|\ =\ d(\rho,\sigma) \nonumber
%\end{eqnarray}
%We have equality if we use the POM $E_i=|v_i><v_i|$.\\
%\end{proof}
%
%In the following we introduce the basic definitions and results for the 
%resolvability theory of a quantum channel $W\colon A\to{\cal H}$.\\

\begin{defi}\rm\label{AchResRate}
1. Let ${\bf W}$ be a quantum channel, let ${\bf E}$ be a measurement process 
on its output space, and let $\varepsilon>0$. We call $R\ge 0$ an
\it $\varepsilon$-achievable resolution rate for ${\bf E}$ \rm if $\,\forall$
processes ${\bf P}$, $\gamma>0$ $\exists\ {\rm process}\ \tilde{\bf P},
n_0\in\mathbb N:$
\begin{equation}\label{arr}
\left(\ \frac{\log_2R(\tilde P^n)}{n}\ <\ R+\gamma \quad\wedge\quad
d_{E^n}(P^nW^n,\tilde P^nW^n)\ <\ \varepsilon\ \right) 
\quad\forall\ n\ge n_0\ .
\end{equation}

2. $R\ge 0$ is an \it achievable resolution rate for ${\bf E}$ 
\index{achievable resolution rate for ${\bf E}$}\rm if
$R$ is an $\varepsilon$-achievable resolution rate for ${\bf E}$ 
for all $\varepsilon>0$.\\

3. Now the channel's \it ($\varepsilon$-)resolution for ${\bf E}$
\index{resolution $S({\bf E})$ for ${\bf E}$}\rm is given as follows:
\begin{eqnarray}\nonumber
S_\varepsilon({\bf E}) &\=& \inf\{ R\ge 0 : \text{R \rm is an
        $\varepsilon$-achievable resolution rate for ${\bf E}$}\},\cr
S({\bf E}) &\=& \inf\{ R\ge 0 : 
   \text{R \rm is an achievable resolution rate for ${\bf E}$}\}.
\end{eqnarray}

4. Fixing in 1.~the input process ${\bf P}$, too, we say that $R\ge 0$ is an 
\it $\varepsilon$-achievable resolution rate for $({\bf P},{\bf E})$ 
\index{achievable resolution rate for $({\bf P},{\bf E})$}\rm if
$\,\forall\,\gamma>0\ \exists$ process $\tilde{\bf P}, n_0\in\mathbb N:$
$$\left(\ \frac{\log_2R(\tilde P^n)}{n}\ <\ R+\gamma \quad\wedge\quad
  d(P^nW^n,\tilde P^nW^n)\ <\ \varepsilon\ \right) \quad\forall\ n\ge n_0\ .$$
Like above we define numbers $S_\varepsilon({\bf P},{\bf E})$ and
$S({\bf P},{\bf E})$.\\
\end{defi}

The following properties are immediate consequences of the definitions:\\

\begin{rem}\rm\label{RateRem}
a) If $\varepsilon\le\varepsilon'$ then 
$S_\varepsilon({\bf E})\ge S_{\varepsilon'}({\bf E})$ and
$S_\varepsilon({\bf P},{\bf E})\ge S_{\varepsilon'}({\bf P},{\bf E})$.\\
b) $S({\bf E}) = \sup_{\varepsilon>0} S_\varepsilon({\bf E})$ and
$S({\bf P},{\bf E}) = \sup_{\varepsilon>0} S_\varepsilon({\bf P},{\bf E})$.\\
c) $S({\bf E})=\sup_{\bf P}S({\bf P},{\bf E})$ and
$S_\varepsilon({\bf E})=\sup_{\bf P}S_\varepsilon({\bf P},{\bf E})$
for all $\varepsilon>0$.\\
\end{rem}

Next we define the notion of uniform resolution rates which will be a useful
tool in the proof of Lemma \ref{loglog<S}.\\

\begin{defi}\rm
Let ${\bf W}$ be a quantum channel, let ${\bf E}$ be a measurement process on 
its output space, and let $\varepsilon>0$. We call $R\ge 0$ a \it 
uniform $\varepsilon$-achievable resolution rate for ${\bf E}$ 
\index{uniform $\varepsilon$-achievable resolution rate for ${\bf E}$}\rm if 
$\,\forall\ \gamma>0\ 
\exists\ n_0\in{\mathbb N}:$
$\forall\ {\rm processes}\ {\bf P}\ \exists\ $process~$\tilde{\bf P}:$
$$\left(\ \frac{\log_2R(\tilde P^n)}{n}\ <\ R+\gamma \quad\wedge\quad
d_{E^n}(P^nW^n,\tilde P^nW^n)\ <\ \varepsilon\ \right) 
\quad\forall\ n\ge n_0\ .$$\smallskip
\end{defi}

\begin{lem}\rm
If $R\ge 0$ is an $\varepsilon$-achievable resolution rate for a
measurement process ${\bf E}$ then $R$ is also a uniform 
$\varepsilon$-achievable resolution rate for ${\bf E}$.\\
\end{lem}

\begin{proof}
Let $R\ge 0$ be an $\varepsilon$-achievable resolution rate for ${\bf E}$, and
let $\gamma>0$. For a fixed process ${\bf P}$ there is a minimum $n_0({\bf P})$
such that for some process $\tilde{\bf P}$:
$$\left(\ \frac{\log_2R(\tilde P^n)}{n}\ <\ R+\gamma \quad\wedge\quad
d_{E^n}(P^nW^n,\tilde P^nW^n)\ <\ \varepsilon\ \right) 
\quad\forall\ n\ge n_0({\bf P})\ .$$
We have to prove that $\sup_{\bf P}n_0({\bf P})<\infty$.\\
Suppose the opposite and let $({\bf P}_k)_k$ be a sequence of processes such
that $n_k\=n_0({\bf P}_k)$ is strictly monotonically increasing (hence
divergent). Define a new process ${\bf P}$ by
$$ P^n \= P_k^n \qquad {\rm for}\ n_{k-1}\le n\le n_k\ .$$
Consider the minimum $k$ for which $n_0({\bf P})<n_k$. Since for
$n_0({\bf P})\le n<n_k$ we have $P^n=P_k^n$ there exist probability 
distributions $\tilde P^n$ for which 
$\frac{\log_2R(\tilde P^n)}{n}\ <\ R+\gamma$ and
$d_{E^n}(P^nW^n,\tilde P^nW^n)\ <\ \varepsilon$. By definition of
$n_0({\bf P}_k)$ there are such $\tilde P^n$ for $n\ge n_k=n_0({\bf P}_k)$,
too. This contradicts $n_0({\bf P}_k)$ being chosen as minimum number.
\end{proof}\bigskip

\begin{lem}\rm\label{loglog<S}
Let ${\bf W}$ be a quantum channel and let ${\bf E}$ be a measurement process 
on its output space. Moreover, let $({\bf P}_k)_{k\in{\mathbb N}}$ be a 
sequence of processes, let $(N_n)_{n\in{\mathbb N}}$ be a sequence of positive
integers, and let $\varepsilon>0$ be with $(\forall\,n\in{\mathbb N})$
$\min_{1\le k < l\le N_n} d_{E^n}(P_k^nW^n,P_l^nW^n) \ge 2\varepsilon$. Then
$\limsup_{n\to\infty} \frac{\log_2\log_2 N_n}{n} \le S_\varepsilon({\bf E})$.\\
\end{lem}

\begin{proof}
By the previous remark, $S_\varepsilon({\bf E})$ is a uniform 
$\varepsilon$-achievable resolution rate for ${\bf E}$.
So, for $\gamma>0$ there is some $n_0\in{\mathbb N}$ and a sequence of 
processes $(\tilde{\bf P}_k)_{k\in{\mathbb N}}$ such that for all 
$n\ge n_0$ and $k\in{\mathbb N}$:
$$\frac{\log_2R(\tilde P_k^n)}{n}
  \ <\ S_\varepsilon({\bf E}) + \gamma \quad\wedge\quad
  d_{E^n}(P_k^nW^n,\tilde P_k^nW^n)\ <\ \varepsilon \ .$$
For fixed $n\in{\mathbb N}$ let's assume that $\tilde P_k^n = \tilde P_l^n$ for
some $1\le k < l\le N_n$. This leads to
$$ d_{E^n}(P_k^nW^n,P_l^nW^n) 
   \ \le\ d_{E^n}(P_k^nW^n,\tilde P_k^nW^n) + d_{E^n}(\tilde P_l^nW^n,P_l^nW^n)
   \ <\ 2\varepsilon \ .$$
Since this is a contradiction it follows that $\tilde P_k^n \not= \tilde P_l^n$
for all $1\le k < l\le N_n$.\\
Now, as there are not more than $|A|^{nM}$ probability distributions on $A^n$ 
of type $M$, and as
for all $n\ge n_0$ and $1\le k\le N_n$ the resolution of $\tilde P_k^n$ is
smaller than $2^{n(S_\varepsilon({\bf E}) + \gamma)} :$
\begin{eqnarray}
N_n  &\le& 
   \sum_{M=1}^{\lfloor 2^{n(S_\varepsilon({\bf E})+\gamma)} \rfloor} |A|^{nM}
   \le\ 2^{n(S_\varepsilon({\bf E})+\gamma)} 
             |A|^{n2^{n(S_\varepsilon({\bf E})+\gamma)}} \nonumber\\
     &=& 2^{n(S_\varepsilon({\bf E})+\gamma)           \nonumber
                     + \log_2{|A|}n2^{n(S_\varepsilon({\bf E})+\gamma)}} \quad
\stackrel{n>\!\!>1}{<}\quad 2^{2^{n(S_\varepsilon({\bf E})+2\gamma)}} \ .
\end{eqnarray}
It follows $\frac{\log_2\log_2 N_n}{n} < S_\varepsilon({\bf E})+2\gamma\ $ 
(for all $\gamma>0$), and
$\limsup_{n\to\infty} \frac{\log_2\log_2 N_n}{n} \le S_\varepsilon({\bf E})$.\\
\end{proof}\bigskip

\begin{thm}\rm\label{C<S}
Let ${\bf W}$ be a quantum channel and let $\lambda_1,\lambda_2>0$ be with
$\lambda_1+\lambda_2<1$. Let $\varepsilon\=1-\lambda_1-\lambda_2$.
There is a measurement process ${\bf E}$ on ${\bf W}$'s output space such that
$\bar C^{sim}(\lambda_1,\lambda_2)\le S_\varepsilon({\bf E})$.\\
\end{thm}

\begin{proof}
For each block length $n\in{\mathbb N}$ let $\{(P_i^n,D_i^n): i=1,\ldots,N\}$ 
be a simultaneous $(n,N,\lambda_1,\lambda_2)$ Q-ID code of maximum size 
$N_n\=N(n,\lambda_1,\lambda_2)$ and ${\bf E}\=(E^n)_{n\in{\mathbb N}}$
with $E^n$ the common refinement of the $D_i^n$ (cf.~Def.~\ref{SimDef}).
Consider a sequence $({\bf P}_i)_{i\in\mathbb N}$ of processes with
${\bf P}_i = (P_i^n)_{n\in\mathbb N}$ where $P_i^n$ is arbitrary for $i>N_n$. 
We have for all $1\le k<l\le N_n$:\footnote{We use that we have for PDs 
$Q,Q'$ on a set $B$: $d_1(Q,Q')=2\sup_{C\subset B}[Q(C)-Q'(C)]$.}
\begin{eqnarray}
 & &\hspace{-1.5cm} d_{E^n}(P_k^nW^n,P_l^nW^n) \nonumber\\
  &=&\  d_1(P_k^nW^n(E^n),P_l^nW^n(E^n)) 
 \ \ge\ d_1(P_k^nW^n(D_k^n),P_l^nW^n(D_k^n))              \nonumber\\
 &\ge&\ 2\,(P_k^nW^n(D_k^n) - P_l^nW^n(D_k^n))
  \ \ge\ 2\,(1-\lambda_1-\lambda_2)\ =\ 2\,\varepsilon \ .\nonumber
\end{eqnarray}
So, Lemma \ref{loglog<S} implies $\bar C(\lambda_1,\lambda_2)
=\limsup_{n\to\infty}\frac{\log_2\log_2 N_n}{n} \le S_\varepsilon({\bf E})$.\\
\end{proof}\bigskip

\begin{rem}\rm\label{Rem4}
If in our definitions we replaced the condition that (\ref{arr}) holds for all
$n\in{\mathbb N}$ by ``(\ref{arr}) holds for an infinite number of 
$n\in{\mathbb N}$,'' there would be no need of the notion of 'uni\-form' 
resolution rates (as the term ``$\exists\,n_0\in{\mathbb N}$'' would disappear
in the definitions). Then Lemma \ref{loglog<S} would state that 
$\liminf_{n\to\infty}\frac{\log_2\log_2 N_n}{n}\le S_\varepsilon({\bf E})$,
and the result of Theorem \ref{C<S} would be 
$C^{sim}(\lambda_1,\lambda_2) \le S_\varepsilon({\bf E})$.\\
\end{rem}

Now we are able to prove the Converse Theorem \ref{ConSim}
up to some fact we deal with in the next section:\\

\begin{proof}
By Theorem \ref{C<S} we have 
$\bar C^{sim}(\lambda_1,\lambda_2)\le S_\varepsilon({\bf E})$. In the following
section we will see (cf.~Theorem \ref{S<C}) that $S_\varepsilon({\bf E})
= \sup_{\bf P}S_\varepsilon({\bf P},{\bf E}) \le \bar C_1$. Analogously we
obtain $C^{sim}(\lambda_1,\lambda_2) \le C_1$ (cf.~the previous remark and
\ref{Rem5}).\\
\end{proof}\bigskip

Theorem \ref{C<S} and Q-ID coding Theorem \ref{DirSim} imply a
\it converse \index{converse resolvability theorem}\rm of 
Resolvability The\-orem \ref{S<C} (cf.~next section):\\

\begin{rem}\rm
Let $0<\varepsilon<1$. There is a measurement process ${\bf E}$ of ${\bf W}$
with $\bar C_0 \le S_\varepsilon({\bf E})$.\\
\end{rem}

The method of upperbounding ID capacities by resolutions immediately leads
to trivial bounds:\\

\begin{rem}\rm\label{TrivCon}
It's easy to see that $S_\varepsilon({\bf E}) \le \log_2 |A|$ always holds,
and $\log_2 |A|$ would remain an upper bound of resolution if
in our definitions we replaced $d_{\bf E}(\rho,\sigma)$ by 
$d(\rho,\sigma) \= \max_{\bf E} d_{\bf E}(\rho,\sigma)$.
This leads to the (natural) bound $C(\lambda_1,\lambda_2) \le \log_2 |A|$.\\
\end{rem}

%-----------------------------------------------------------------------------
\bigskip
%\clearpage\pagestyle{empty}\cleardoublepage\pagestyle{plain}
\section[Transmission Capacity as Upper Bound of Resolution]{Transmission Capacity as Upper Bound of\\ Resolution}
%------------------------------------------------------
\bigskip

This section is devoted to the proof of the following theorem. Like in the 
previous section we will do this following the ideas of \cite{han93}.\\

\begin{thm}\rm\label{S<C}
Let $({\bf P},{\bf E})$ be a pair of processes for ${\bf W}$, and let
$\varepsilon>0$.
Then $S_\varepsilon({\bf P},{\bf E}) \le \bar C_1$.\\
\end{thm}

We introduce some basic concepts to prove this theorem:\\

\begin{defi}\rm\label{lip}
Let ${\bf A}=(A_n)_{n\in{\mathbb N}}$ be a sequence of random variables. Its
\it limsup in probability \index{limsup in probability}\rm is the number
$$ \bar{\bf A}\ \=\ \min \{ \beta\in\bar{\mathbb R} \,|\,
   \forall\,\varepsilon>0:\lim_{n\to\infty}P[A_n\ge\beta+\varepsilon]=0\}\ .$$
\smallskip\end{defi}

\begin{defi}\rm
Let $({\bf P},{\bf E})$ be a pair of processes for ${\bf W}$. For every
$n\in{\mathbb N}$ let
$$ P_{(P^n,E^n)}(x^n,y) \ \=\ P^n(x^n) W^n_{x^n}(E^n_y) $$
be the joint distribution of the classical channel that outputs the result of
measurement $E^n$ on $W^n$'s output. Let
$$ i_{(P^n,E^n)}(x^n,y) 
  \ \=\ \log_2 \frac{W^n_{x^n}(E^n_y)} {P^nW^n(E^n_y)} $$
be its \it information density. 
\index{information density}\rm The \it sup-information rate 
\index{sup-information rate}\rm
$\bar{\bf I}({\bf P},{\bf E})$ is defined to be the limsup in probability of
the normalized information density 
$I_n \= \frac{1}{n}\,i_{(P^n,E^n)}$.\\
\end{defi}

\begin{lem}\rm\label{I<C}
Let $({\bf P},{\bf E})$ be a pair of processes for ${\bf W}$. Then
$\bar{\bf I}({\bf P},{\bf E}) \le \bar C_1$.\\
\end{lem}

\begin{proof}
Let's assume that this is false and that there is a pair of processes 
$({\bf P},{\bf E})$ such that for some $\alpha,\gamma>0:$
$$ P_{(P^n,E^n)}
   \left[\frac{1}{n}\,i_{(P^n,E^n)} > \bar C_1+\gamma\right] > \alpha$$
for infinitely many integers $n$. With this assumption we will be able to 
construct for those integers -- if only they are large enough -- an
$(n,M_n,1-\frac{\alpha}{3})$ Q code, with $M_n$ some integer fulfilling
$$ \bar C_1+\frac{\gamma}{3}\ \le\ \frac{\log_2 M_n}{n}\ \le\ \bar C_1+\frac{\gamma}{2}\ .$$
As the first inequality contradicts to the definition of $\bar C_1$ the 
lemma will be proved.\\
Given $E^n=(E^n_1,\ldots,E^n_{b_n})$ let for
every $x^n\in A^n$
\begin{eqnarray}
D(x^n) & \= & \left\lbrace y\in\{1,\ldots,b_n\} : 
                \frac{1}{n}\,i_{(P^n,E^n)}(x^n,y) 
                > \bar C_1+\gamma \right\rbrace ,\nonumber\\
{\rm and\ then}\hspace{1.5cm}& &\rightline{}\nonumber\\
     G & \= & \left\lbrace x^n\in A^n : 
                W^n_{x^n}(\sum_{y\in D(x^n)}E^n_y) 
                \ge \frac{\alpha}{2} \right\rbrace.\nonumber
\end{eqnarray}
We choose the codewords $c_i\in G$ successively by the random selection method
with probability
$$ Q^n(x^n)\ \=\ 
\begin{cases}
\frac{P^n(x^n)}{P^n(G)}, & \text{if $x^n\in G$,}\\
                      0, & \text{otherwise.}
\end{cases}   $$
%\cases{\frac{P^n(x^n)}{P^n(G)}, & if $x^n\in G$,\cr
%                                            0, & otherwise.\cr}   
The decoding operator of the codeword $c_i$ is defined to be the operator
$\sum_{y\in D_i}E^n_y$ where
$ D_i \= D(c_i)\, \setminus\, \bigcup_{j<i} D(c_j)\,.$\\
The success probability is
\begin{eqnarray}
W^n_{c_i}(\sum_{y\in D_i}E^n_y)
 & \ge & W^n_{c_i}(\sum_{y\in D(c_i)}E^n_y)
       - \sum_{j<i} W^n_{c_i}(\sum_{y\in D(c_j)}E^n_y) \nonumber\\
 & \ge & \frac{\alpha}{2}
       - \sum_{j<i} W^n_{c_i}(\sum_{y\in D(c_j)}E^n_y)\ .\nonumber
\end{eqnarray}
For the expected value of the last summands
\begin{eqnarray}
{\mathbb E}_{Q^n}\,W^n_{c_i}(\sum_{y\in D(c_j)}E^n_y)
 & \le & \frac{1}{P^n(G)}
\sum_{x^n\in A^n}P^n(x^n)W^n_{x^n}(\sum_{y\in D(c_j)}E^n_y) \nonumber\\
 & = & \frac{1}{P^n(G)} P^nW^n(\sum_{y\in D(c_j)}E^n_y) ,\nonumber
\end{eqnarray}
holds, where
$$ P^nW^n(\sum_{y\in D(c_j)}E^n_y)
 \ =\ \sum_{y=1}^{b_n} P^nW^n(E^n_y)
     \cdot 1_{\{ y\in D(c_j) \}}
 \ <\ 2^{-n(\bar C_1+\gamma)}\ . $$
Here the inequality holds because
\begin{eqnarray}
y\in D(c_j) \quad &\Leftrightarrow& \quad \frac{1}{n}\log_2
       \frac{W^n_{c_j}(E^n_y)}{P^nW^n(E^n_y)}\ >\ \bar C_1+\gamma \nonumber\\
 &\Leftrightarrow& \quad P^nW^n(E^n_y)\ <\ W^n_{c_j}(E^n_y)
      \cdot 2^{-n(\bar C_1+\gamma)} .\nonumber
\end{eqnarray}
So, we get for the expected success probability
$$ {\mathbb E}_{Q^n}\, W^n_{c_i}(\sum_{y\in D_i}E^n_y)
 \ \ge\ \frac{\alpha}{2} - \frac{1}{P^n(G)}\, M_n \cdot 2^{-n(\bar C_1+\gamma)}
 \ \ge\ \frac{\alpha}{2} - \frac{1}{P^n(G)}\, 2^{-n\frac{\gamma}{2}}\ .$$
For some $\zeta\ge 0$ and random variable $Z<1$ clearly
$ {\mathbb E}_P\,Z \le P[Z\ge\zeta] + \zeta $ holds. Therefore:
\begin{eqnarray}
P^n(G) & = & P^n\left[W^n_{x^n}(\sum_{y\in D(x^n)}E^n_y)
                   \ge \frac{\alpha}{2}\right]
\ \ge\ {\mathbb E}_{P^n}\ W^n_{x^n}(\sum_{y\in D(x^n)}E^n_y)
     - \frac{\alpha}{2} \nonumber\\
& = & \sum_{x^n\in A^n} P^n(x^n)
      W^n_{x^n}(\sum_{y\in D(x^n)}E^n_y) - \frac{\alpha}{2} \nonumber\\
& = & P_{(P^n,E^n)} \left[\frac{1}{n}\,i_{(P^n,E^n)}>\bar C_1+\gamma\right]
      - \frac{\alpha}{2} \nonumber\\
& \ge & \alpha - \frac{\alpha}{2}\qquad (\text{... by assumption}). \nonumber
\end{eqnarray}
Hence
$$ {\mathbb E}_{Q^n}\, W^n_{c_i}(\sum_{y\in D_i}E^n_y)
 \ \ge\ \frac{\alpha}{2} - \frac{2}{\alpha}\, 2^{-n\frac{\delta}{2}}
 \ \ge\ \frac{\alpha}{3} $$
-- if only $n$ is large enough --, and there is certainly one codeword 
$c_i\in G$ with the desired success probability.
\end{proof}\bigskip

\begin{lem}\rm[cf.~\cite{han93}, p.~758]
Let $Q$ and $R$ be probability distributions on a finite set. Then for
every $\mu>0$:
$$d_1(Q,R)\ \le\ \frac{2}{\log_2e}\mu+2Q\left[\log_2\frac{Q}{R}>\mu\right]\ .$$\\
\end{lem}

\begin{lem}\rm\label{S<I}
Let $({\bf P},{\bf E})$ be a pair of processes for ${\bf W}$, and let 
$\varepsilon>0$.
Then $S_\varepsilon({\bf P},{\bf E}) \le \bar{\bf I}({\bf P},{\bf E})$.\\
\end{lem}

\begin{proof}
Let $\gamma>0$. We show that by the random selection method there is a
process $\tilde{\bf P}$ such that
$$ \frac{\log_2 R(\tilde P^n)}{n}\ \le\ \bar{\bf I}({\bf P},{\bf E}) + \gamma
   \qquad {\rm and} \qquad    \lim_{n\to\infty} 
   d_{E^n} (P^nW^n,\tilde P^nW^n)\ =\ 0\ .$$
This works as follows:
For fixed $n$ let 
$M\=\lfloor 2^{n(\bar{\bf I}({\bf P},{\bf E})+\gamma)} \rfloor$.
Each $M$-tuple $(c_1,\ldots,c_M)\in (A^n)^M$ of codewords gives rise to the
$M$-type probability distribution
$$ \tilde P^n_{(c_1,\ldots,c_M)}(x^n)
  \ \=\ \frac{1}{M} \sum_{i=1}^M {\bf 1}_{\{x^n=c_i\}}\ .$$
We will show that
$$ \lim_{n\to\infty} {\mathbb E}_{P^n}\ d_{E^n}
   (P^nW^n,\tilde P^n_{(c_1,\ldots,c_M)}W^n)\ =\ 0\ ,$$
interpreting the $c_1,\ldots,c_M$ as independent outcomes of a random 
experiment with underlying probability distribution $P^n$. This directly 
implies our claim.\\
Recall that $d_{E^n}
   (P^nW^n,\tilde P^n_{(c_1,\ldots,c_M)}W^n)=$ $ d_1(P^nW^n(E^n),
         \tilde P^n_{(c_1,\ldots,c_M)}W^n(E^n))$ (cf. Def. \ref{distance}), and
by the previous lemma it is enough to show that for every $\mu>0$ the 
following expression goes to $0$ as $n$ tends to infinity:
\begin{eqnarray}
& &\hspace{-1.5cm}
   \sum_{c_1\in A^n}\cdots\sum_{c_M\in A^n}P^n(c_1)\cdots P^n(c_M)\nonumber\\
& &\hspace{1.5cm}
   \sum_{y=1}^{b_n}\tilde P^n_{(c_1,\ldots,c_M)}W^n(E^n_y)
  \cdot{\bf 1}{\{\log_2
    \frac{\tilde P^n_{(c_1,\ldots,c_M)}W^n(E^n_y)}
         {P^nW^n(E^n_y)} > \mu\} } \nonumber\\
& &\hspace{-1.5cm}=\ \frac{1}{M}\sum_{j=1}^M 
     \sum_{c_1\in A^n}\cdots\sum_{c_M\in A^n}P^n(c_1)\cdots P^n(c_M)\nonumber\\
& &\hspace{1.5cm}\sum_{y=1}^{b_n}W^n_{c_j}(E^n_y)\cdot{\bf 1}{\{\log_2
    \frac{\tilde P^n_{(c_1,\ldots,c_M)}W^n(E^n_y)}
         {P^nW^n(E^n_y)} > \mu\} }\ . \nonumber
\end{eqnarray}
Since all $M$ summands are equal this is just 
\begin{eqnarray}
& &\hspace{-1.2cm}
   \sum_{c_2\in A^n}\cdots\sum_{c_M\in A^n}P^n(c_2)\cdots P^n(c_M) \nonumber\\
& &\hspace{1.2cm}
   \sum_{c_1\in A^n}\sum_{y=1}^{b_n}P_{(P^n,E^n)}(c_1,y)\cdot{\bf 1}{\{\log_2
    \frac{\tilde P^n_{(c_1,\ldots,c_M)}W^n(E^n_y)}
         {P^nW^n(E^n_y)} > \mu\} } \nonumber\\
& &\hspace{-1.2cm} \le\ P_{(P^n,E^n)}\left[\frac{1}{M}2^{i_{(P^n,E^n)}}
                                   > \tau \right] \label{sum}\\
& &\hspace{1.2cm} 
   +\ {\mathbb E}_{P^nW^n(E^n)} P_{(P^n)^{M-1}}\left[\frac{1}{M}
        \sum_{j=2}^M2^{i_{(P^n,E^n)}(c_j,y)} > 1+\tau\right]\ .\nonumber
\end{eqnarray}
Here $\tau\=\frac{1}{2}(2^\mu-1)>0$, and the last inequality holds because
\begin{eqnarray}
&\ &\hspace{-2cm} {\bf 1}{\{\log_2
    \frac{\tilde P^n_{(c_1,\ldots,c_M)}W^n(E^n_y)}
         {P^nW^n(E^n_y)} > \mu\} }
 \ =\ {\bf 1}{\{\log_2
    \frac{\frac{1}{M}\sum_{j=1}^M W^n_{c_j}(E^n_y)}
         {P^nW^n(E^n_y)} > \mu\} }\nonumber\\
 &=& {\bf 1}{\{\log_2
    \frac{1}{M}\sum_{j=1}^M 2^{i_{(P^n,E^n)}(c_j,y)} > \mu\} }\nonumber\\
 &=& {\bf 1}{\{\frac{1}{M}2^{i_{(P^n,E^n)}(c_1,y)}
   +\frac{1}{M}\sum_{j=2}^M 2^{i_{(P^n,E^n)}(c_j,y)} > 1+2\tau\} }\nonumber\\
 &\le& {\bf 1}{\{\frac{1}{M}2^{i_{(P^n,E^n)}(c_1,y)} > \tau\} }
    + {\bf 1}{\{
    \frac{1}{M}\sum_{j=2}^M 2^{i_{(P^n,E^n)}(c_j,y)} > 1+\tau\} }\ .
    \nonumber
\end{eqnarray}
The first summand of (\ref{sum}) is easy to handle:
$$ P_{(P^n,E^n)}\left[\frac{1}{M}2^{i_{(P^n,E^n)}}
                                   > \tau \right]
  \ =\ P_{(P^n,E^n)}\left[\frac{1}{n}i_{(P^n,E^n)}
          > \frac{\log_2\tau}{n} + \frac{\log_2M}{n} \right] $$
$$ \ \le\ P_{(P^n,E^n)}\left[I_n
  > \frac{\log_2\tau}{n} +\bar{\bf I}({\bf P},{\bf E}) +\frac{\gamma}{2}\right]
 \quad \underset{n\to\infty}{\longrightarrow} \quad 0\ .$$
For the last inequality (which holds for large $n$) recall that
$M=\lfloor 2^{n(\bar{\bf I}({\bf P},{\bf E})+\gamma)} \rfloor$.\\
Now, since for every $y\in\{1,\ldots,b_n\}$
\begin{eqnarray}
%& & \hspace{-2.1cm}
{\mathbb E}_{(P^n)^{M-1}} 
       \frac{1}{M}\sum_{j=2}^M2^{i_{(P^n,E^n)}(c_j,y)}
 & = & \frac{1}{M}\sum_{j=2}^M {\mathbb E}_{P^n}\, 
                 2^{i_{(P^n,E^n)}(c_j,y)} \nonumber\\
 \le\ {\mathbb E}_{P^n}\, 2^{i_{(P^n,E^n)}(c_2,y)}
 & = & \sum_{c_2\in A^n} P^n(c_2) \frac{W^n_{c_2}(E^n_y)}
                                      {P^nW^n(E^n_y)}
 \ =\ 1\ ,\nonumber
\end{eqnarray}
we can apply Chebychev's Inequality:
\begin{eqnarray}
& &\hspace{-2.5cm}
   P_{(P^n)^{M-1}} \left[\frac{1}{M}\sum_{j=2}^M2^{i_{(P^n,E^n)}(c_j,y)}
                      > 1+\tau \right] \nonumber\\
 & \le & \frac{1}{\tau^2}\,{\rm Var}_{(P^n)^{M-1}}\left[\frac{1}{M}
         \sum_{j=2}^M2^{i_{(P^n,E^n)}(c_j,y)}\right] \nonumber\\
 & = & \frac{1}{\tau^2}\, \frac{1}{M^2} \sum_{j=2}^M {\rm Var}_{P^n} 
      \left[2^{i_{(P^n,E^n)}(c_j,y)}\right] \nonumber\\
 & \le & \frac{1}{\tau^2}\, \frac{1}{M}\, {\rm Var}_{P^n}
      \left[2^{i_{(P^n,E^n)}(c_2,y)}\right] \nonumber\\
 & \le & \frac{1}{\tau^2}\, \frac{1}{M}\, {\mathbb E}_{P^n}
      \left[(2^{i_{(P^n,E^n)}(c_2,y)})^2\right]\ .\nonumber
\end{eqnarray}
And we get the following upper bound for the second summand of (\ref{sum}):
\begin{eqnarray}
&\ &\hspace{-1.8cm} \frac{1}{\tau^2}\,\frac{1}{M}\, {\mathbb E}_{P^nW^n(E^n)}
  {\mathbb E}_{P^n} \left[(2^{i_{(P^n,E^n)}(c_2,y)})^2\right] \nonumber\\
  &=& \frac{1}{\tau^2}\, \frac{1}{M} \sum_{c_2\in A^n}\sum_{y=1}^{b_n}
       P^n(c_2)P^nW^n(E^n_y) \left(
 \frac{W^n_{c_2}(E^n_y)}{P^nW^n(E^n_y)}
 \right)^2 \nonumber\\
 & = & \frac{1}{\tau^2}\, {\mathbb E}_{(P^n,E^n)}
       \left[\frac{1}{M}\,2^{i_{(P^n,E^n)}(c_2,y)}\right] \nonumber\\
 & = & \frac{1}{\tau^2} \Big( {\mathbb E}_{(P^n,E^n)}
       \left[\frac{1}{M}\,2^{i_{(P^n,E^n)}(c_2,y)}\cdot
             {\bf 1}\{ \frac{1}{M}\,2^{i_{(P^n,E^n)}(c_2,y)}
                       \le 2^{-n\frac{\gamma}{2}} \} \right] \nonumber\\
 & & \qquad +\ {\mathbb E}_{(P^n,E^n)}
       \left[\frac{1}{M}\,2^{i_{(P^n,E^n)}(c_2,y)}\cdot
             {\bf 1}\{ \frac{1}{M}\,2^{i_{(P^n,E^n)}(c_2,y)}
                       > 2^{-n\frac{\gamma}{2}} \} \right] \Big) \nonumber\\
 & \le & \frac{1}{\tau^2} \left( 2^{-n\frac{\gamma}{2}}
         + P_{(P^n,E^n)} \left[
           \frac{1}{M}\,2^{i_{(P^n,E^n)}} > 2^{-n\frac{\gamma}{2}}
           \right] \right) \nonumber\\
 & \underset{n\gg 1}{\le} & \frac{1}{\tau^2} \left( 2^{-n\frac{\gamma}{2}}
        + P_{(P^n,E^n)} \left[
          I_n > \bar{\bf I}({\bf P},{\bf E}) + \frac{\gamma}{3} \right] \right)
    \quad \underset{n\to\infty}{\longrightarrow} \quad 0 \nonumber
\end{eqnarray}
\end{proof}\smallskip

\begin{rem}\rm\label{Rem5}
If in Definition \ref{lip} we replaced the limit by a liminf, Lemma \ref{I<C}
could be formulated as $\bar{\bf I}({\bf P},{\bf E}) \le C_1$, and with the
changes proposed by Remark \ref{Rem4}, Lemma \ref{S<I} would still hold. Thus
Theorem \ref{S<C} would state that $S_\varepsilon({\bf P},{\bf E}) \le C_1$
(with changed definitions for $S_\varepsilon({\bf P},{\bf E})$).\\
\end{rem}

%---------------------------------------------------------------------------

\section{Acknowledgments}
%-------------------------

I'd like to thank my colleague Andreas Winter for the discussion on topics
of this paper as well as for some ideas he gave me about how to look at
several aspects of quantum information theory. 
I'd also like to thank Prof. Rudolf Ahlswede who introduced me to the area of 
information theory and who proposed to me the problem of identification by
the use of quantum channels.\\

%-----------------------------------------------------------------------------
\bigskip
%\clearpage\pagestyle{empty}\cleardoublepage\pagestyle{plain}

%-----------------------------------------------------------------------------
\bigskip
%\clearpage\pagestyle{empty}\cleardoublepage\pagestyle{plain}

%\addcontentsline{toc}{section}{\numberline{}Index}
%\input{dis.ind}

\end{document}